\documentclass[letterpaper, 10 pt, conference]{ieeeconf}

\usepackage{amsmath}
\usepackage{mathtools}
\usepackage{amssymb}
\usepackage{amsfonts}
\usepackage{graphicx}
\usepackage{booktabs}
\usepackage{multirow}
\usepackage{balance}
\newtheorem{theorem}{Theorem}
\newtheorem{proposition}{Proposition}

\IEEEoverridecommandlockouts

\title{\LARGE \bf
Predictive Safety Filter for Multi-Modal Learned Policies
}

\author{Myeonggeun Jun and Eunhyek Joa$^{*}$%
\thanks{This research was supported by KOSEF/SNU-IAMD.
This work was supported by the New Faculty Startup Fund from Seoul National University.
This paper was supported by Korea Institute for Advancement of Technology (KIAT) grant funded by the Korea Government (MOTIE) (HRD Program for Industrial Innovation).
This work was conducted with the support of the
\(\ulcorner\)Gyeonggi-do Joint R\&D Project in Autonomous Driving\(\lrcorner\)
managed by the Advanced Institute of Convergence Technology with financial
resources from Gyeonggi-do in year 2025.
This work is supported by the Korea Agency for Infrastructure Technology Advancement (KAIA) grant funded by the Ministry of Land, Infrastructure and Transport (Grant code RS-2021-KA162182).}%
\thanks{M. Jun and E. Joa are with Vehicle Autonomy and Intelligence Lab, Department of Mechanical Engineering, Seoul National University, Republic of Korea. $^{*}$Corresponding author: Eunhyek Joa ({\tt eunhyekj@snu.ac.kr}).}%
}
\begin{document}

\maketitle
\begingroup
\renewcommand{\thefootnote}{}
\footnotetext{\scriptsize \copyright~2026 IEEE. Personal use of this material is permitted. Permission from IEEE must be obtained for all other uses, in any current or future media, including reprinting/republishing this material for advertising or promotional purposes, creating new collective works, for resale or redistribution to servers or lists, or reuse of any copyrighted component of this work in other works.}
\addtocounter{footnote}{-1}
\endgroup
\thispagestyle{empty}
\pagestyle{empty}

\begin{abstract}
This paper presents a predictive safety filter (PSF) for learned policies with multi-modal outputs. 
We represent the multi-modal output as a Gaussian mixture model (GMM) and seek a trajectory that remains in a high-likelihood region while satisfying state and input constraints for nonlinear systems. 
The resulting problem is formulated as a Kullback-Leibler (KL) divergence minimization between the filter output and the GMM. 
By modeling the filter output as a Gaussian with vanishing covariance, this objective reduces in the zero-variance limit to a log-sum-exp cost, yielding a tractable nonlinear program (NLP) solved once per time step. 
We show that the performance degradation due to safety enforcement is bounded by a function of the  worst-case KL divergence. 
The proposed method is validated in  autonomous parking scenarios and compared with two predictive safety filter baselines: one that filters only the maximum-likelihood mode and one that filters all candidate modes and selects the best filtered one according to the  finite-horizon discounted return.
The proposed method achieves higher aggregate goal-reaching success than both baselines while requiring only a single NLP per step and  achieving lower wall-clock time than Parallel SF in the full-mixture comparison.
\end{abstract}


\section{Introduction}

Data-driven policies have been demonstrated to solve complex motion planning and control tasks~\cite{fisac2019general}. 
Their deployment in real-world safety-critical systems, however, requires mechanisms that guarantee safety, since unsafe behavior may lead to property damage, loss of trust in autonomous systems, and human harm. 
Safety filters address this challenge by acting as a modular certification layer that minimally modifies an unsafe reference input to enforce safety and operational constraints~\cite{wabersich2024ddsf}.

Safety filters can be broadly categorized into three classes. 
First, Hamilton--Jacobi (HJ) reachability~\cite{bansal2017hj,fisac2019general} computes safe sets through backward reachability analysis. While it provides worst-case safety guarantees, it suffers from the curse of dimensionality in high-dimensional systems.
Second, control barrier functions (CBFs)~\cite{ames2017cbf,ames2019cbf} enforce forward invariance through pointwise quadratic programs. 
Although computationally efficient, they require the explicit construction of a barrier certificate, which can be challenging for complex constraint sets. 
Third, predictive safety filters (PSFs)~\cite{wabersich2018linear,wabersich2021predictive} solve a receding-horizon optimal control problem at each time step, thereby incorporating trajectory-level constraints directly into the filtering problem.

A common PSF formulation minimizes the deviation between a single reference trajectory and the filtered output~\cite{tearle2021predictive, didier2021adaptive, leeman2023psf}. This formulation implicitly assumes that the planner provides a single reference maneuver at each time step and that this reference remains temporally consistent over the prediction horizon.
While these methods provide strong safety guarantees, they do not account for the possibility that the planner may consider multiple maneuver candidates simultaneously.
In practice, however, data-driven policies often produce multimodal output distributions~\cite{salzmann2020trajectron, schmerling2018multimodal}, such as Gaussian mixture model (GMM) policies that assign probability mass to multiple candidate modes.
For example, a trajectory prediction module may output several plausible future paths with associated likelihoods, or a  reinforcement learning (RL) policy may maintain a mixture distribution over distinct strategies.
In such cases, the preferred mode may vary over time, and committing to a single reference trajectory, even the one with the highest likelihood at a given time step, may fail to capture the full structure of the planner output.
This can in turn lead to conservative behavior or abrupt mode switching when multiple modes have comparable scores~\cite{gray2013predictive}.

In this paper, we extend the predictive safety filter framework to compute a safe trajectory that minimally deviates from a multi-modal reference distribution. 
Existing methods typically consider either a unimodal reference distribution or a single trajectory generated by a data-driven policy, whereas the proposed method explicitly accounts for multimodal reference trajectories. 
Specifically, we model the multi-modal output as a GMM over the state-trajectory space, where each mode represents a distinct candidate maneuver. The safety filter then computes a trajectory that minimizes the Kullback--Leibler (KL) divergence~\cite{kullback1951information} between the filter output and the multimodal distribution induced by the data-driven policy while satisfying all state and input constraints.

The contributions of this paper are as follows:
\begin{itemize}
\item We formulate a predictive safety filter for multi-modal learned policies based on KL-divergence minimization, yielding a single nonlinear program (NLP) at each time step.
\item We show a performance--safety trade-off bound that quantifies the return loss induced by safety filtering via Pinsker's inequality and the Performance Difference Lemma (Theorem~\ref{thm:tradeoff}).
\item With numerical validation in  100 autonomous parking scenarios under two obstacle-layout conditions, the proposed algorithm improves aggregate goal-reaching success over existing safety filters: one that filters only the maximum-likelihood mode and one that filters all modes and selects the best one (Sections~\ref{sec:setup}--\ref{sec:results}).
\end{itemize}

\section{Problem Setup}

\subsection{System Dynamics}

Consider a nonlinear discrete-time system of the form
\begin{equation}
x_{i+1|k} = F\!\left(x_{i|k},\; u_{i|k}\right),
\label{eq:dynamics}
\end{equation}
where $x_{i|k} \in \mathcal{X} \subset \mathbb{R}^{n_x}$ is the predicted state, $u_{i|k} \in \mathcal{U} \subset \mathbb{R}^{n_u}$ is the planned input at time step~$k+i$, both computed at time step~$k$, $N$ is the prediction horizon, and $i = 0,\dots,N{-}1$.
The initial condition is $x_{0|k} = x_k$.
The predicted trajectory must satisfy
\begin{equation}
x_{i|k} \in \mathcal{X}, \quad u_{i|k} \in \mathcal{U}, \quad \forall\, i,
\label{eq:constraints}
\end{equation}
where $\mathcal{X}$ and $\mathcal{U}$ encode all state and input constraints, including collision avoidance and actuator limits.

\subsection{GMM Prior over Trajectories}

We assume access to a learned policy that produces $M$ candidate trajectory modes, each representing a distinct maneuver strategy.
This multi-modal output is encoded as a  GMM over the full state trajectory space $\mathbb{R}^{n_x(N+1)}$:
\begin{equation}
P_{\mathrm{GMM}}(x_{0:N|k})
  = \sum_{m=1}^{M} \alpha_m\,
    \mathcal{N}\!\left(x_{0:N|k};\;\mu_m,\;\Sigma_m\right),
\label{eq:gmm}
\end{equation}
where $\alpha_m>0$, $\sum_{m}\alpha_m=1$,
$\mu_m\!\in\!\mathbb{R}^{n_x(N+1)}$ is the mean trajectory of mode~$m$,
and $\Sigma_m\!\in\!\mathbb{R}^{n_x(N+1)\times n_x(N+1)}$ is the covariance.

\subsection{Predictive Safety Filter}
The safety filter seeks a trajectory close to the prior~$P_{\mathrm{GMM}}$ while satisfying constraints~\eqref{eq:constraints}.
Specifically, we want to compute a solution to the following nonlinear optimization problem:
\begin{align}
\min_{\{x_{i|k},\,u_{i|k}\}} \;\;
  & D_{KL}\!\Big(Q\big(x_{0:N|k}\big)\;\Big\|\;P_{\mathrm{GMM}}\big(x_{0:N|k}\big)\Big) \nonumber\\[4pt]
\text{s.t.}\;\;
  & x_{0|k} = x_k, \label{eq:obj_kl}\\
  & x_{i+1|k} = F(x_{i|k},\,u_{i|k}), \nonumber\\
  & x_{i|k}\in\mathcal{X},\quad u_{i|k}\in\mathcal{U}, \nonumber\\
  & i=0,\dots,N{-}1, \nonumber
\end{align}
where $Q$ is the distribution induced by the filter's trajectory choice.
Note that the objective function is $D_{KL}(Q\|P)$ instead of $D_{KL}(P\|Q)$.
Minimizing $D_{KL}(Q\|P)$ penalizes placing mass where $P$ is small, thereby enforcing that the selected trajectory lies in a region of high probability density.
This is the \emph{mode-seeking} direction, which naturally avoids trajectories that the planner considers unlikely.
In contrast, the reverse direction $D_{KL}(P\|Q)$ would encourage the filter to cover all modes, which is not related in selecting a single feasible trajectory.

\section{Tractable Reformulation of \eqref{eq:obj_kl}}
\label{sec:approach}
The main challenge in solving \eqref{eq:obj_kl} is that evaluating and minimizing the KL divergence with respect to the GMM is generally intractable when $Q$ is an arbitrary distribution.  

This section derives a tractable reformulation of the KL divergence objective~\eqref{eq:obj_kl}.
Since the safety filter outputs a single trajectory $\hat{x}_{0:N|k}$, we model its output distribution as a Gaussian with vanishing covariance:
\begin{equation}
Q(x) = \mathcal{N}\!\left(x;\; \hat{x}_{0:N|k},\; \varepsilon I\right), \quad \varepsilon > 0.
\label{eq:q_eps}
\end{equation}

Next, we show that KL divergence in \eqref{eq:obj_kl} is equivalent to a negative log likelihood (NLL) with \eqref{eq:q_eps}.
\begin{proposition}[Negative Log Likelihood Equivalence]
\label{prop:eps_independence}
Let $\mathcal{F}$ be a compact feasible set.
As $\varepsilon \to 0$,
\begin{equation}
\arg\min_{\hat{x} \in \mathcal{F}}\;D_{KL}(Q \| P_{\mathrm{GMM}})
  \;\to\; \arg\min_{\hat{x} \in \mathcal{F}}\;\bigl[-\log P_{\mathrm{GMM}}(\hat{x})\bigr].
\label{eq:wellposed}
\end{equation}
\end{proposition}

\begin{proof}
The KL divergence decomposes as~\cite{cover2006elements}
\[
D_{KL}(Q \| P) = -H(Q) - \mathbb{E}_{Q}\!\left[\log P(x)\right].
\]
where the entropy
\[
H(Q) = \tfrac{n}{2}\log(2\pi e\,\varepsilon), \quad n = n_x(N{+}1),
\]
depends on~$\varepsilon$ and is constant in~$\hat{x}$.
For the cross-entropy, Taylor-expanding to second order gives
\begin{equation}
\mathbb{E}_{Q}\!\left[f(x)\right]
  = f(\hat{x})
    + \tfrac{\varepsilon}{2}\,\mathrm{tr}\!\bigl(\nabla^2 f(\hat{x})\bigr)
    + O(\varepsilon^2).
\label{eq:taylor}
\end{equation}
where $f = \log P_{\mathrm{GMM}}$. Under $\Sigma_m \succ 0$, $f$ is smooth.
The $O(\varepsilon)$ correction is bounded uniformly over the compact set~$\mathcal{F}$ and vanishes as $\varepsilon \to 0$, leaving
\[
\arg\min_{\hat{x} \in \mathcal{F}}\;\bigl[-\log P_{\mathrm{GMM}}(\hat{x})\bigr].\]
\end{proof}

Substituting the GMM~\eqref{eq:gmm} into the NLL from Proposition~\ref{prop:eps_independence} yields the NLP solved at every step~$k$:
\begin{align}
\min_{\substack{x_{i|k},\,u_{i|k}}} \;\;
  & -\log P_{\mathrm{GMM}}(x_{0:N|k}) \nonumber\\[4pt]
\text{s.t.}\;\;
  & x_{0|k} = x_k, \label{eq:nlp}\\
  & x_{i+1|k} = F(x_{i|k},\,u_{i|k}), \nonumber\\
  & x_{i|k}\in\mathcal{X},\quad u_{i|k}\in\mathcal{U}, \nonumber\\
  & i=0,\dots,N{-}1. \nonumber
\end{align}
The NLP~\eqref{eq:nlp} is solved once per time step via an interior-point method.
The first element $u_{0|k}^{*}$ of the optimal input sequence is applied to the plant, and the procedure is repeated at step $k{+}1$ in a receding-horizon fashion.
Only a \emph{single} NLP is solved per time step, regardless of the number of modes~$M$.

\smallskip
\noindent\textit{Remark.}
In practice, regularization terms such as control effort $\sum_i u_{i|k}^{\!\top} R\, u_{i|k}$ or input rate penalties may be appended to the objective in~\eqref{eq:nlp}.
The performance bound described in Theorem \ref{thm:tradeoff} remains valid as it depends on the distributional shift, not on the NLP cost function.
The NLP is non-convex; warm-starting aids local convergence but does not guarantee global optimality.

\section{Properties}

\subsection{Feasible Mode Selection}

A conventional PSF projects a single reference trajectory onto the feasible set by minimizing the deviation between the reference and the filtered output~\cite{tearle2021predictive}.
While this guarantees the feasibility, it remains tied to the selected reference mode even when that mode can no longer complete the task under the constraints.
As a result, the filtered trajectory may remain safe but become overly conservative or fail to make task progress.
In contrast, the proposed safety filter in \eqref{eq:nlp} evaluates all modes simultaneously and can redirect the solution toward another mode that is better suited to completing the task.
The log-sum-exp objective combines all modes: weights determine preference and covariances their influence range, favoring feasible high-likelihood trajectories under the full mixture.

Let $\mathcal{F}$ denote the feasible trajectory set of~\eqref{eq:nlp}.
Define the squared Mahalanobis distance from a trajectory~$x_{0:N}$ to mode~$m$ as
\begin{equation}
d_m(x_{0:N}) := (x_{0:N} - \mu_m)^{\!\top}\Sigma_m^{-1}(x_{0:N} - \mu_m).
\label{eq:mahal}
\end{equation}
Let
\begin{align*}
x_{0:N}^{*} &:= \arg\min_{x_{0:N} \in \mathcal{F}} \bigl[-\log P_{\mathrm{GMM}}(x_{0:N})\bigr], \\
x_{0:N,m}^{*} &:= \arg\min_{x_{0:N} \in \mathcal{F}} d_m(x_{0:N}).
\end{align*}

\begin{theorem}[Feasible Mode Selection]
\label{thm:mode_selection}
Suppose $\mathcal{F} \neq \emptyset$.
The GMM prior density evaluated at the multi-modal optimum~$x_{0:N}^{*}$ is at least as large as the highest single-mode density at its own optimizer:
\begin{equation}
P_{\mathrm{GMM}}(x_{0:N}^{*})
  \;\geq\; \max_{m=1,\dots,M}\;
  \alpha_m\,\mathcal{N}\!\left(x_{0:N,m}^{*};\;\mu_m,\;\Sigma_m\right).
\label{eq:nll_bound}
\end{equation}
\end{theorem}

\begin{proof}
Since negative logarithm is a monotonically decreasing function, the maximizer of $P_{\mathrm{GMM}}(x_{0:N})$ over the feasible set $\mathcal{F}$ is $x_{0:N}^{*}$, which is the minimizer of the $-\log P_{\mathrm{GMM}}(x_{0:N})$ over $\mathcal{F}$.

Moreover, by the definition of the Gaussian distribution, $x_{0:N,m}^{*}$ is the maximizer of $\max_{x_{0:N} \in \mathcal{F}}\;
  \alpha_m\,\mathcal{N}\!\left(x_{0:N};\;\mu_m,\;\Sigma_m\right)$ for the given $m$.

Remind that the GMM is a sum of non-negative components. Thus,
\begin{align*}
P_{\mathrm{GMM}}(x_{0:N})
  &= \sum_{m=1}^{M}\alpha_m\,\mathcal{N}(x_{0:N};\mu_m,\Sigma_m) \\
  &\geq\; \alpha_{m'}\,\mathcal{N}(x_{0:N};\mu_{m'},\Sigma_{m'})
\end{align*}
for any $m'$.

Therefore, maximizing over $\mathcal{F}$ on both sides and then over $m'$ gives~\eqref{eq:nll_bound}.
\end{proof}

Theorem~\ref{thm:mode_selection} shows that the full-mixture objective can favor another feasible mode when the maximum-likelihood mode becomes infeasible, without explicit switching.

\subsection{Performance--Safety Trade-off}
Let $\pi$ denote the nominal policy and $J(\pi) := \mathbb{E}_{\pi}\!\left[\sum_{t=0}^{\infty}\gamma^{t}\,r_t\right]$ the expected discounted return under discount factor $\gamma \in (0,1)$. When the safety filter~\eqref{eq:nlp} operates in a receding-horizon fashion, the resulting closed-loop policy $\tilde{\pi}$ differs from~$\pi$ due to constraint enforcement.
We assume the worst-case distributional shift is bounded by
\noindent
\begin{equation}
\begin{aligned}
D_{KL}^{\max}\!\bigl(\tilde{\pi}\,\big\|\,\pi\bigr)
&:= \sup_{x\in\mathcal{X}_{\mathrm{feas}}}
D_{KL}\!\bigl(\tilde{\pi}(\cdot|x)\,\big\|\,\pi(\cdot|x)\bigr) \leq \bar{\delta},\quad\text{}
\end{aligned}
\label{eq:kl_bound}
\end{equation}

where $\mathcal{X}_{\mathrm{feas}}$ is the feasible state set and $\bar{\delta}<\infty$ is a uniform bound.

\begin{theorem}[Performance--Safety Trade-off]
\label{thm:tradeoff}
Let $\pi$ be the nominal policy with expected return $J(\pi)$, and $\tilde{\pi}$ be the safety-filtered policy satisfying~\eqref{eq:kl_bound}.
Let $\epsilon := \|A^{\pi}\|_{\infty}$ where $A^{\pi}$ is the advantage function of~$\pi$.
Then
\begin{equation}
J(\tilde{\pi}) \geq J(\pi) - \frac{\sqrt{2}\,\gamma\epsilon}{(1{-}\gamma)^2}\,\sqrt{\bar{\delta}}.
\label{eq:tradeoff}
\end{equation}
\end{theorem}

\smallskip
\begin{proof}
By Pinsker's inequality~\cite{tsybakov2009introduction},
\noindent
\begin{equation}
D_{TV}^{\max}\!\bigl(\tilde{\pi}\,\big\|\,\pi\bigr)
  \leq \sqrt{\tfrac{1}{2}\,D_{KL}^{\max}\!\bigl(\tilde{\pi}\,\big\|\,\pi\bigr)}
  \leq \sqrt{\tfrac{\bar{\delta}}{2}}.
\quad\text{}
\label{eq:pinsker}
\end{equation}
By the Performance Difference Lemma~\cite{kakade2002approximately},
\begin{equation}
\bigl|J(\tilde{\pi}) - J(\pi)\bigr|
  \leq \frac{2\gamma\epsilon}{(1{-}\gamma)^2}\,
  D_{TV}^{\max}\!\bigl(\tilde{\pi}\,\big\|\,\pi\bigr)
  \quad\text{}.
\label{eq:pdl}
\end{equation}
Substituting~\eqref{eq:pinsker} into~\eqref{eq:pdl} gives
\[
J(\pi) - J(\tilde{\pi})
  \leq \frac{2\gamma\epsilon}{(1{-}\gamma)^2}\,\sqrt{\tfrac{\bar{\delta}}{2}}
  = \frac{\sqrt{2}\,\gamma\epsilon}{(1{-}\gamma)^2}\,\sqrt{\bar{\delta}},
\]
which yields~\eqref{eq:tradeoff}.
\end{proof}

\section{Simulation Setup}
\label{sec:setup}

The proposed safety filter is evaluated in a simulated autonomous parking scenario with static obstacles.
The NLP is formulated using CasADi~\cite{andersson2019casadi} and solved with IPOPT~\cite{wachter2006implementation}.

\subsection{Vehicle Model}

The vehicle is modeled by a kinematic bicycle model with state $x = [X, Y, \psi, v]^\top$ (position, heading, speed) and control input $u = [\delta, v_{\mathrm{cmd}}]^\top$ (steering angle, speed command).
The discrete-time dynamics with time step $\Delta t = 0.1\,\mathrm{s}$ are
\begin{align}
X_{i+1} &= X_i + v_i \cos(\psi_i)\,\Delta t, \nonumber\\
Y_{i+1} &= Y_i + v_i \sin(\psi_i)\,\Delta t, \nonumber\\
\psi_{i+1} &= \psi_i + \tfrac{v_i}{L}\tan(\delta_i)\,\Delta t, \label{eq:bicycle}\\
v_{i+1} &= v_{\mathrm{cmd},i}, \nonumber
\end{align}
where $L = 2.0\,\mathrm{m}$ is the wheelbase.
The constraint sets are
\[
\mathcal{U} = \bigl\{u : |\delta| \le 1.0\,\mathrm{rad},\; v_{\mathrm{cmd}} \in [0,\,3.0]\,\mathrm{m/s}\bigr\},
\]
and $\mathcal{X}$ encodes the collision-free region.
Collision avoidance is enforced via point-mass constraints with inflated obstacles.
For obstacle~$j$ with center $(o_j^x, o_j^y)$ and half-dimensions $(h_j^x, h_j^y)$, the vehicle center $(X_i, Y_i)$ must satisfy
\begin{equation}
\max\!\bigl(|X_i - o_j^x| - h_j^x,\; |Y_i - o_j^y| - h_j^y\bigr) \geq d_{\mathrm{safe}},
\label{eq:collision}
\end{equation}
where $d_{\mathrm{safe}} > 0$ accounts for the vehicle footprint and an additional safety margin.
 IPOPT solves the resulting nonsmooth NLP to local optimality.
All reported results are local solutions.

\subsection{Environment}
\label{sec:environment}

The parking environment is a $30 \times 20\,\mathrm{m}$ rectangular area bounded by walls.
Four goal positions are placed at $(10, -5)$, $(10, 5)$, $(-10, -5)$, and $(-10, 5)$.

We evaluate two obstacle-layout conditions. In the dominant-mode-infeasible (blocking-wall) condition, an additional right-side wall blocks the initially dominant modes, and six interior obstacles are position-perturbed from the base layout. In the dominant-mode-feasible (open-layout) condition, the blocking wall is removed and eight obstacles are independently position-perturbed from a symmetric template, leaving the dominant side accessible. Obstacle dimensions are fixed in both conditions. This pairing tests mode redirection when the dominant mode is infeasible and performance when it remains feasible.
The vehicle footprint is $1.5 \times 0.8\,\mathrm{m}$.

\subsection{GMM from Learned Policy}

The GMM prior is constructed by combining four independently trained goal-conditioned soft actor-critic (SAC) policies, one for each parking goal. We refer to their weighted multi-modal composition as the combined SAC policy. It outputs $M=4$ trajectory modes corresponding to distinct parking approach directions (right-down, right-up, left-down, left-up) with associated weights $\alpha_m$.
The mode means $\mu_m$ are generated by forward-propagating the learned actions through the kinematic bicycle model~\eqref{eq:bicycle} over the prediction horizon~$N$; the policy is reevaluated at each predicted state while keeping the mode identity fixed.
Let $p(x)=[X,Y]^\top$, $h(x)=[\cos\psi,\sin\psi]^\top$, and let $g_m$ be the goal associated with mode~$m$. The reported experiments use the state-dependent scores and weights
\begin{equation}
\begin{aligned}
d_m^{(q)}(x)
  &=\|p(x)-g_m\|_q,\\
\rho_m^{(q)}(x)
  &=\frac{\lambda_d}{d_m^{(q)}(x)+1}
    +\lambda_\psi h(x)^\top
    \frac{g_m-p(x)}{\|g_m-p(x)\|_2},\\
\alpha_m^{(q)}(x)
  &=\frac{\exp\!\left(\rho_m^{(q)}(x)\right)}
  {\sum_{j=1}^{M}\exp\!\left(\rho_j^{(q)}(x)\right)},
  \\
q&\in\{1,2,\infty\} .
\end{aligned}
\label{eq:distance_weights}
\end{equation}
The main comparison uses $q=2$, $\lambda_d=2$, and $\lambda_\psi=0.7$; Table~\ref{tab:norm_ablation} varies only $q$. The heading term favors modes aligned with the current vehicle yaw, while the reciprocal-distance term retains goal proximity and the softmax preserves all positive mixture weights.
The covariances $\Sigma_m$ are set to  time-factorized position covariance blocks propagated from policy variances with positive diagonal regularization; this is a practical approximation rather than an exact uncertainty propagation through the nonlinear dynamics.
 Each goal-conditioned SAC policy is trained for $2 \times 10^5$ environment steps.
The reward function encourages reaching the  assigned goal while penalizing collisions.

\subsection{Safety Filter Parameters}

 All safety-filter comparisons use matched prediction horizons $N\in\{10,20\}$, corresponding to $1.0$ and $2.0\,\mathrm{s}$.
 Control-effort and speed-tracking terms are included as NLP regularization.
\section{Simulation Results}
\label{sec:results}

 Four configurations are compared in the  same scenario set:
\begin{itemize}
\item \textbf{No-SF}: Unfiltered output of the  combined SAC policy.
\item \textbf{ Single-mode SF}: A single-mode predictive safety filter that tracks the highest-weight mode via a quadratic cost, with $M=1$ and matched $N\in\{10,20\}$.
\item \textbf{Parallel SF}: A parallel safety filtering approach inspired by batch trajectory optimization~\cite{adajania2022multi}. Each GMM mode is filtered independently, and the feasible trajectory with the highest finite-horizon discounted return is executed. This requires $M$ concurrent NLP solves per step.
\item \textbf{Proposed}: The log-sum-exp safety filter~\eqref{eq:nlp}.
\end{itemize}

\subsection{Qualitative Analysis}

\begin{figure*}[t]
\centering
\includegraphics[width=\textwidth]{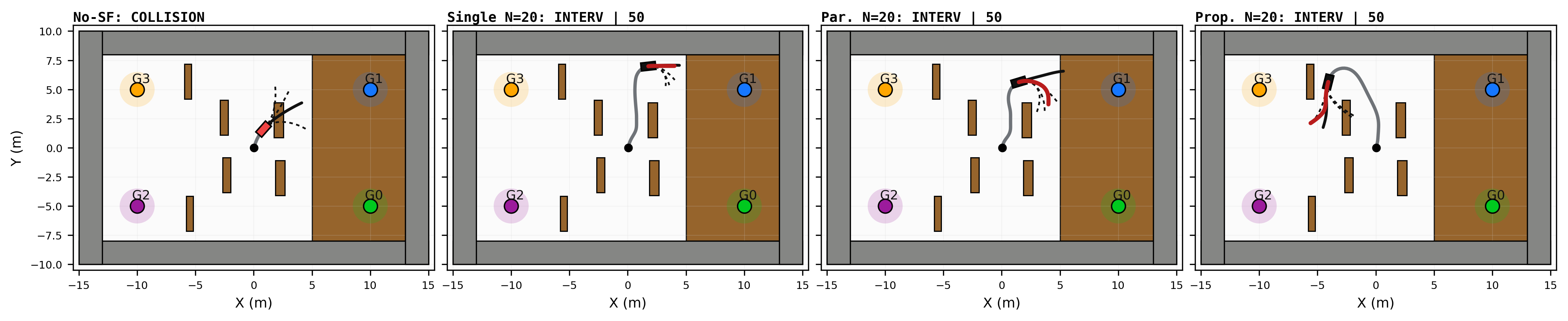}
\vspace{-8mm}
\caption{Intermediate  states ($t \approx 5\,\mathrm{s}$) in a representative blocking-wall episode. No-SF has collided; all safety filters use $N=20$.}
\label{fig:parallel_intermediate}
\label{fig:sf_mid}
\includegraphics[width=\textwidth]{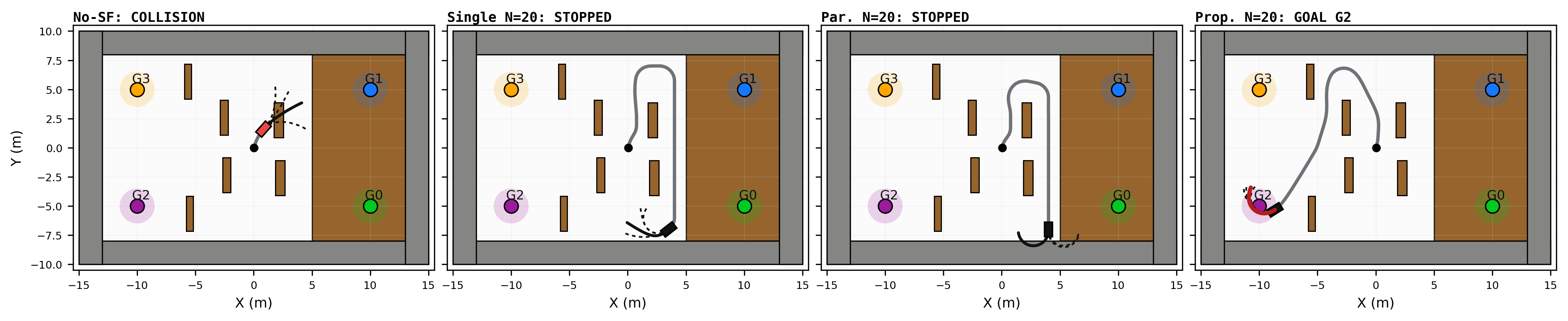}
\vspace{-8mm}
\caption{Terminal  states in the same episode. No-SF collides;  Single-mode SF and Parallel SF stop; Proposed reaches G2.}
\label{fig:parallel_terminal}
\label{fig:sf_end}
\end{figure*}

\begin{figure*}[t]
\centering
\includegraphics[width=\textwidth]{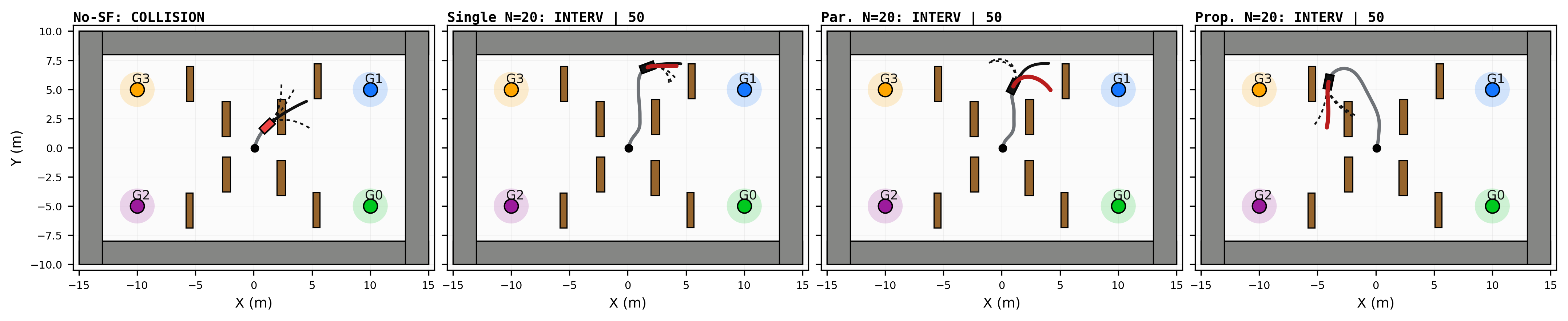}
\vspace{-8mm}
\caption{Intermediate states ($t \approx 5\,\mathrm{s}$) in a representative symmetric open-layout episode. No-SF has collided; all safety filters use $N=20$.}
\label{fig:sf_open_mid}
\includegraphics[width=\textwidth]{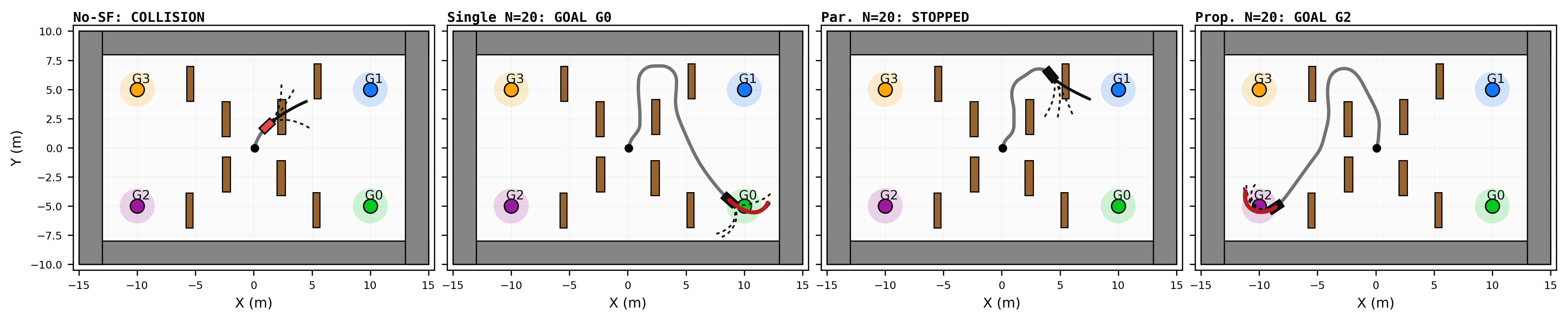}
\vspace{-8mm}
\caption{Terminal states in the same symmetric open-layout episode. No-SF collides; Single-mode SF reaches G0, Parallel SF stops, and Proposed reaches G2.}
\label{fig:sf_open_end}
\vspace{2mm}
\end{figure*}

Fig.~\ref{fig:sf_mid}--\ref{fig:sf_end}  compare No-SF, Single-mode SF, Parallel SF, and Proposed in a representative blocking-wall episode at $N=20$. Fig.~\ref{fig:sf_open_mid}--\ref{fig:sf_open_end} show the corresponding intermediate and terminal comparisons in the symmetric open layout.

At $t \approx 5\,\mathrm{s}$ (Fig.~\ref{fig:sf_mid}), No-SF follows the initially dominant right-side mode and collides. Single-mode SF remains feasible by stopping near the wall, while Parallel SF remains tied to independently filtered references and also stops. Proposed instead optimizes the full log-sum-exp objective and redirects toward a feasible left-side mode without an explicit switching rule. At the terminal state (Fig.~\ref{fig:sf_end}), Proposed reaches G2 while the two baselines remain stopped, illustrating the feasible-mode redirection characterized by Theorem~\ref{thm:mode_selection}.

Gray denotes the executed trajectory, black the GMM rollouts, and red the safety-filter output.
In the symmetric open layout, Single-mode SF reaches G0 and Proposed reaches G2, while Parallel SF stops (Fig.~\ref{fig:sf_open_mid}--\ref{fig:sf_open_end}).

\subsection{Quantitative Analysis}

All configurations are evaluated over  100 fixed scenarios shared by all methods, arranged as 50 matched pairs using the two conditions defined in Section~\ref{sec:environment}.
Each pair shares its start state.
Success is defined as reaching any goal region (distance $< 1.5\,\mathrm{m}$) within $T=50\,\mathrm{s}$ without collision.

\begin{table}[t]
\centering
\caption{
Performance at matched horizons $N\in\{10,20\}$ under two paired conditions (50 scenarios each). Single-mode SF uses $M=1$; Parallel SF and Proposed use $M=4$. Continuous values are mean $\pm$ sample SD.}
\label{tab:quantitative}
\scriptsize
\setlength{\tabcolsep}{3pt}
\begin{tabular*}{\columnwidth}{@{\extracolsep{\fill}}clccc@{}}
\toprule
$N$ & Method & Succ. & Coll. & $J(\tilde{\pi})$ \\
\midrule
\multicolumn{5}{l}{\textit{Dominant-mode-infeasible}, $J(\pi)=528.25\pm1.83$} \\
10 & Single-mode SF & 0/50 & 0/50 & $186.15\pm68.39$ \\
10 & Parallel SF & 0/50 & 0/50 & $242.90\pm4.75$ \\
10 & Proposed & \textbf{48/50} & 0/50 & \textbf{\boldmath$325.71\pm49.40$} \\
\midrule
20 & Single-mode SF & 0/50 & 0/50 & $213.11\pm3.21$ \\
20 & Parallel SF & 0/50 & 0/50 & $271.48\pm20.64$ \\
20 & Proposed & \textbf{50/50} & 0/50 & \textbf{\boldmath$365.67\pm33.26$} \\
\midrule
\multicolumn{5}{l}{\textit{Dominant-mode-feasible}, $J(\pi)=528.25\pm1.83$} \\
10 & Single-mode SF & 0/50 & 0/50 & $188.91\pm55.16$ \\
10 & Parallel SF & 0/50 & 0/50 & $249.45\pm5.66$ \\
10 & Proposed & \textbf{50/50} & 0/50 & \textbf{\boldmath$337.83\pm17.48$} \\
\midrule
20 & Single-mode SF & \textbf{50/50} & 0/50 & \textbf{\boldmath$382.23\pm2.30$} \\
20 & Parallel SF & 17/50 & 0/50 & $345.22\pm46.93$ \\
20 & Proposed & 49/50 & 0/50 & $360.13\pm42.53$ \\
\bottomrule
\end{tabular*}
\end{table}

\paragraph{Theorem~\ref{thm:mode_selection}: Feasible Mode Selection}
Theorem~\ref{thm:mode_selection} shows that the multi-modal filter  favors a feasible high-probability trajectory from the given GMM distribution.
The empirical results in Table~\ref{tab:quantitative} illustrate this property. Proposed has the highest aggregate success at both horizons: 98/100 at $N=10$ and 99/100 at $N=20$, with 0/50 collisions in every condition.
Parallel SF selects by finite-horizon discounted return, which does not capture the longer-horizon consequences represented by the combined SAC value function; an infeasible mode-specific reference can therefore lead to stopping.
Single-mode SF's 50/50 open-layout result at $N=20$ is horizon-specific: the corresponding $N=10$ runs stall, whereas the longer horizon enables a transition toward G0. Its blocking-wall result remains 0/50 because it cannot redirect from the infeasible dominant side. Proposed's sole $N=20$ miss is due to a late-stage stall near an obstacle boundary.

\paragraph{Theorem~\ref{thm:tradeoff}: Performance--Safety Trade-off}
To illustrate the empirical performance trade-off consistent with Theorem~\ref{thm:tradeoff}, we report the nominal and safety-filtered returns.
The nominal return $J(\pi)$ is the average  discounted return of the  combined SAC policy over  the 50 matched initial states in the \emph{obstacle-free} environment.
The filtered return $J(\tilde{\pi})$ is measured under each safety filter configuration and reported as mean $\pm$ sample standard deviation.
Relative to the nominal return $J(\pi)$, Proposed yields the closest $J(\tilde{\pi})$ in both blocking-wall cases and at $N=10$ in the open layout; Single-mode SF is closest in the open layout at $N=20$. Thus, Proposed improves aggregate completion and infeasible-mode redirection without uniformly maximizing return.

\subsection{Computational Performance}
\label{sec:comp}

 Table~\ref{tab:scaling} reports per-step wall-clock time on an Intel Core i9-14900KS desktop with 48\,GB RAM under WSL2. Each configuration uses eight concurrent scenario workers with numerical-library threads fixed to one; Parallel SF additionally executes $M$ mode-specific NLPs concurrently. The wall-clock measurement includes rollout generation and optimization. Proposed's NLP dimension is independent of $M$, and at $M=4$ it is $2.9\times$ and $4.3\times$ faster than Parallel SF in the two conditions.

\begin{table}[t]
\centering
\caption{Per-step wall-clock time at $N=20$ across mode counts (50 scenarios per condition; ms, mean $\pm$ sample SD).}
\label{tab:scaling}
\scriptsize
\renewcommand{\arraystretch}{1.02}
\setlength{\tabcolsep}{1.5pt}
\begin{tabular*}{\columnwidth}{@{\extracolsep{\fill}}ccccc@{}}
\toprule
& \multicolumn{2}{c}{\textit{Dominant-mode-infeasible}} & \multicolumn{2}{c}{\textit{Dominant-mode-feasible}} \\
\cmidrule(lr){2-3}\cmidrule(lr){4-5}
$M$ & Parallel SF & Proposed & Parallel SF & Proposed \\
\midrule
1 & $60.4\pm3.2$ & \textbf{\boldmath$74.2\pm7.0$} & $195.4\pm23.5$ & \textbf{\boldmath$127.5\pm50.9$} \\
3 & $187.1\pm62.2$ & \textbf{\boldmath$77.6\pm2.7$} & $206.3\pm103.3$ & \textbf{\boldmath$80.9\pm3.8$} \\
4 & $236.8\pm18.7$ & \textbf{\boldmath$80.4\pm4.5$} & $351.4\pm250.4$ & \textbf{\boldmath$81.4\pm7.1$} \\
\bottomrule
\end{tabular*}
\end{table}

Table~\ref{tab:norm_ablation} replaces $L_2$ in~\eqref{eq:distance_weights} with $L_1$ and $L_\infty$ at $M=4$, $N=20$. All three norms reach 50/50 goals in the dominant-mode-infeasible condition, while open-layout success differs by only one episode. Their similar returns and 0/50 collisions indicate limited sensitivity to the distance norm.

\begin{table}[t]
\centering
\caption{Proposed weight-norm sensitivity at $M=4$ and $N=20$ (50 scenarios per condition; mean $\pm$ sample SD).}
\label{tab:norm_ablation}
\scriptsize
\renewcommand{\arraystretch}{1.02}
\setlength{\tabcolsep}{1.5pt}
\begin{tabular*}{\columnwidth}{@{\extracolsep{\fill}}lcccc@{}}
\toprule
& \multicolumn{2}{c}{\textit{Dominant-mode-infeasible}} & \multicolumn{2}{c}{\textit{Dominant-mode-feasible}} \\
\cmidrule(lr){2-3}\cmidrule(lr){4-5}
Norm & Success & Return & Success & Return \\
\midrule
$L_1$ & 50/50 & $365.71\pm33.76$ & 50/50 & $361.23\pm38.02$ \\
$L_2$ & 50/50 & $365.67\pm33.26$ & 49/50 & $360.13\pm42.53$ \\
$L_\infty$ & 50/50 & $364.67\pm35.39$ & 49/50 & $360.24\pm41.54$ \\
\bottomrule
\end{tabular*}
\end{table}

\section{Conclusion}

This paper presented a predictive safety filter that incorporates a GMM trajectory prior through KL-divergence minimization and solves one log-sum-exp NLP per step.
The full-mixture solution is at least as compatible with the prior as any feasible single-mode solution (Theorem~\ref{thm:mode_selection}), and the return degradation is bounded by $O(\!\sqrt{\bar{\delta}})$ (Theorem~\ref{thm:tradeoff}).
In the representative blocking-wall episode, Proposed redirected to a feasible goal while  Single-mode SF stalled.
Across the evaluated conditions, Proposed achieved the highest aggregate completion rate, maintained zero collisions, and required lower per-step wall-clock time than Parallel SF in the full-mixture comparison. Performance was also largely insensitive to the choice of weight norm.
Results are limited to the tested weighting and covariance model; local convergence and mixture composition can still affect completion.
The current formulation does not guarantee recursive feasibility.
Future work will investigate  terminal-set or backup-policy designs, systematic sensitivity to mixture weights and covariances, and extensions to dynamic obstacles.

\balance

\end{document}